\documentclass[12pt]{article}
\usepackage{graphicx}
\usepackage{siunitx}
\usepackage{lineno}
\usepackage{multirow}%
\usepackage{xspace}
\usepackage{adjustbox}
\usepackage{url} 
\usepackage{doi}
\usepackage{hyperref} 

\hypersetup{
	pdftitle={Inclusive and differential cross section measurements of ttbb production in the 
	lepton+jets channel at 13 TeV},
	pdfsubject={TOP2023 Conference Proceedings},
	pdfauthor={Emanuel Pfeffer},
	pdfkeywords={CERN LHC CMS ttbb differential}
}

\newcommand\pubnumber{CMS-CR-2024-003}
\newcommand\pubdate{\today}

\newcommand{\ttjets}{\ensuremath{\ttbar{+}\textrm{jets}}\xspace}
\newcommand{\NA}{\textsc{N/A}\xspace}
\newcommand{\HT}{{\ensuremath{H_{\mathrm{T}}}}\xspace}

\newcommand{\ttbar}{\ensuremath{\text{t}\bar{\text{t}}}\xspace}

\newcommand{\ttbb}{\ensuremath{\text{t}\bar{\text{t}}\text{b}\bar{\text{b}}}\xspace}
\newcommand{\tttt}{\ensuremath{\text{t}\bar{\text{t}}\text{t}\bar{\text{t}}}\xspace}
\newcommand{\ttHbb}{t\ensuremath{\bar{\text{t}}\text{H}(\text{b}{\bar{\text{b}}})}\xspace}

\newcommand{\PQb}{\ensuremath{\mathrm{b}}}
\newcommand{\PQt}{\ensuremath{\mathrm{t}}}

\newcommand{\Pg}{\ensuremath{\mathrm{g}}}

 \newcommand{\PAQb}{\ensuremath{\mathrm{\overline{b}}}}

\newcommand{\PAQt}{\ensuremath{\mathrm{\overline{t}}}}

\newcommand{\POWHEG}{\textsc{Powheg}\xspace}

\newcommand{\POWHEGBOX}{{\POWHEG}\textsc{-box-res}\xspace}

\newcommand{\sjfbLONG}{${\geq}6$ jets: ${\geq}4$b\xspace}
\newcommand{\fjtbLONG}{${\geq}5$ jets: ${\geq}3$b\xspace}
\newcommand{\sjtbtlLONG}{${\geq}6$ jets: ${\geq}3$b, ${\geq}3$ light\xspace}
\newcommand{\sejfbtlLONG}{${\geq}7$ jets: ${\geq}4$b, ${\geq}3$ light\xspace}

\newcommand{\ttbbPP}{\textsc{Powheg+ol+p8} \ttbb \textsc{4fs}\xspace}
\newcommand{\ttbarPP}{\textsc{Powheg+p8} \ttbar \textsc{5fs}\xspace}
\newcommand{\ttbarPH}{\textsc{Powheg+h7} \ttbar \textsc{5fs}\xspace}

\newcommand{\ttbbSherpa}{\textsc{sherpa+ol} \ttbb \textsc{4fs}\xspace}

\newcommand{\mT}{\ensuremath{m_{\mathrm{T}}}\xspace}
\newcommand{\mtop}{\ensuremath{m_{\PQt}}\xspace}
\newcommand{\mbottom}{\ensuremath{m_{\PQb}}\xspace}
\newcommand{\mTtop}{\ensuremath{m_{\mathrm{T},\PQt}}\xspace}
\newcommand{\mTi}{\ensuremath{m_{\mathrm{T},i}}\xspace}
\newcommand{\pT}{\ensuremath{p_{\mathrm{T}}}\xspace}
\newcommand{\pTi}{\ensuremath{p_{\mathrm{T},i}}\xspace}
\newcommand{\sumttbbg}{\ensuremath{\sum_{i=\PQt,\PAQt,\PQb,\PAQb,\Pg}}\xspace}
\newcommand{\prodttbb}{\ensuremath{\prod_{i=\PQt,\PAQt,\PQb,\PAQb}}\xspace}
\newcommand{\hdamp}{\ensuremath{\mathrm{\textit{h}}_{\mathrm{damp}}}\xspace}
\newcommand{\muR}{\ensuremath{\mu_{\mathrm{R}}}\xspace}
\newcommand{\muF}{\ensuremath{\mu_{\mathrm{F}}}\xspace}

\newcommand{\HERWIG} {{\textsc{herwig}}\xspace}

\newcommand{\MADGRAPH} {\textsc{MadGraph}\xspace}
\newcommand{\MCATNLO} {\textsc{mc@nlo}\xspace}

\newcommand{\PYTHIA} {{\textsc{pythia}}\xspace}
\newcommand{\SHERPA} {{\textsc{sherpa}}\xspace}

\newcommand{\MGvATNLO}{\MADGRAPH{}5\_a\MCATNLO}

\newcommand{\GeV}{\ensuremath{\,\text{Ge\hspace{-.08em}V}}\xspace}

\newcommand{\HTl}{\ensuremath{H^{\text{light}}_{\mathrm{T}}}\xspace}

\newcommand{\DR}{\ensuremath{\Delta R}\xspace}
\newcommand{\bb}{\ensuremath{\PQb\PQb}\xspace}
\newcommand{\deltaR}{\ensuremath{\DR}\xspace}
\newcommand{\deltaRX}[1]{\ensuremath{\deltaR_{#1}}\xspace}

\newcommand{\dRbb}{\deltaRX{\bb}}

\newcommand{\dRbbavg}{\ensuremath{\deltaR_{\bb}^{\text{avg}}}\xspace}

\newcommand{\bbextra}{{\ensuremath{\bb^{\text{extra}}}}\xspace}
\newcommand{\dRbbextra}{\ensuremath{\deltaR(\bbextra)}\xspace}
\newcommand{\mbbextra}{\ensuremath{m(\bbextra)}\xspace}

\def\institute{Institute for Experimental Particle Physics\\
Karlsruhe Institute of Technology, Germany}
\def\authemail{\footnote{Contact: emanuel.pfeffer@cern.ch}}

\def\Title#1{\begin{center} {\Large #1 } \end{center}}
\def\Author#1{\begin{center}{ \sc #1} \end{center}}
\def\Address#1{\begin{center}{ \it #1} \end{center}}

\newcommand\pubblock{\rightline{\begin{tabular}{l} \pubnumber\\
         \pubdate  \end{tabular}}}
\newenvironment{Abstract}{\begin{quotation}  }{\end{quotation}}
\newenvironment{Presented}{\begin{quotation} \begin{center} 
             PRESENTED AT\end{center}\bigskip 
      \begin{center}\begin{large}}{\end{large}\end{center} \end{quotation}}
\def\Acknowledgements{\bigskip  \bigskip \begin{center} \begin{large}
             \bf ACKNOWLEDGEMENTS \end{large}\end{center}}

\def\beq{\begin{equation}}
\def\eeq#1{\label{#1}\end{equation}}
\def\eeqn{\end{equation}}

\def\beqa{\begin{eqnarray}}
\def\eeqa#1{\label{#1}\end{eqnarray}}
\def\eeqan{\end{eqnarray}}

\let\bar=\overbar

\def\Dslash{\not{\hbox{\kern-4pt $D$}}}
\def\dslash{\not{\hbox{\kern-2pt $\del$}}}

\def\msb{{\bar{\ssstyle M \kern -1pt S}}}

\usepackage{xspace}

\begin{document}
\begin{titlepage}
\pubblock

\vfill
\Title{Inclusive and differential cross section measurements of \ttbb production in the lepton+jets channel 
at $\sqrt{s} = \SI{13}{TeV}$}
\vfill 
\Author{Emanuel Pfeffer\authemail\\
on behalf of the CMS Collaboration}
\Address{\institute}
\vfill
\begin{Abstract}
Differential cross section measurements of the associated production of top quark 
and b quark pairs, \ttbb, are presented.
The results are based on data from proton-proton collisions collected by the CMS detector at the 
LHC at a centre-of-mass energy of $\sqrt{s} = \SI{13}{TeV}$, corresponding to an integrated  
luminosity of $\SI{138}{fb^{-1}}$. Four fiducial phase space regions are defined targeting distinct 
aspects of the \ttbb process.
Kinematic variables are defined at the stable particle level and distributions are unfolded to the particle 
level through maximum likelihood fits. 
The cross sections are measured in the lepton+jets decay channel of the top quark pair, using events 
with exactly one isolated electron or muon and at least five jets.
The results are compared with predictions from several event generators. 
The differential measurements have relative uncertainties in the range of $2\!-\!50\%$, 
depending on the phase space and the observable.
\end{Abstract}
\vfill
\begin{Presented}
$16^\mathrm{th}$ International Workshop on Top Quark Physics\\
(Top2023), 24--29 September, 2023
\end{Presented}
\vfill
\end{titlepage}
\def\thefootnote{\fnsymbol{footnote}}
\setcounter{footnote}{0}

\section{Introduction}
The two widely separated energy scales of the top and the bottom quark, each with a very different, 
non-negligible mass, result in particular complexity when it comes to modeling processes in which both 
quark flavours occur simultaneously as pairs~\cite{Buccioni:2019plc, Jezo:2018yaf}.
These processes arise in proton-proton collisions at 
the CERN LHC as associated production of top and bottom quark-antiquark pairs, \ttbb. 
Due to the different emerging energy scales, the measurement of these processes is a major test of 
perturbative quantum chromodynamics (QCD) calculations.
In addition, a precise knowledge of \ttbb production is crucial in order to be able to understand other 
important processes for which \ttbb forms the leading irreducible background. 
Among these processes and searches are the measurement of the associated production of top quark 
pairs with Higgs bosons with the Higgs boson decaying into a pair of b quarks 
(\ttHbb)~\cite{ATLAS:tthbb2022, CMS:2020grm}, and 
measurements of the simultaneous production of four top quarks (\tttt)~\cite{ATLAS:2023ajo, 
CMS:TOP-22-013}. 
Both processes directly access a key parameter of the Standard Model (SM), the top quark 
Yukawa coupling. 

\section{The CMS detector}
 The Compact Muon Solenoid (CMS) detector is an experiment at the CERN LHC with a 
 superconducting solenoid reaching a high magnetic field of $\SI{3.8}{T}$. An all-silicon pixel and 
 strip 
 tracker is positioned as close as possible to the beam axis. The trackers are surrounded by a 
 lead-tungstate 
 scintillating-crystals electromagnetic calorimeter, and followed by a brass-scintillator sampling 
 hadron 
 calorimeter. All these components are located inside the solenoid. Outside the magnet the iron yoke of the 
 flux-return is combined with muon detectors. 
 This onion-like construction allows almost every spatial angle to be covered so that as much of the 
 collision  
 as possible may be reconstructed and recorded.
 A two-staged trigger system is employed for the recording in order to be able to handle the resulting data 
 volume of over 40 million collisions per second. 
 Only after the calculation by a processor farm of the second tier it is possible to store the collected 
 data~\cite{TheCMSCollaboration_2008}.

\section{Modelling approaches}
In the analysis, six different modelling approaches for \ttbb production are compared with the 
measurements.
These approaches differ in the Monte Carlo~(MC) event generators used, the processes simulated 
at the matrix-element (ME) level, the applied parton showering (PS) models, the scales used and 
additional parameters which can be found in~\cite{cmscollaboration2023inclusive}.
The main difference between the simulations, apart from the settings mentioned before, is the simulated 
process at ME level.
Most importantly, a distinction can be made between models using a \ttbar simulation at the ME level 
at next-to-leading order (NLO) accuracy in QCD and models using \ttbb at the ME level at NLO.
Two of the tested models describe \ttbar at the ME level, three others test \ttbb at ME level and one 
simulation describes \ttjets with up to two additional jets at ME level. 
All the simulation approaches are summarized in Table~\ref{tab:generatorsettings}.

\begin{table}[!ht]
	\centering
	\caption{Generator settings for all considered modeling approaches of \ttbb production.
		The top quark mass value is set to $\mtop = 172.5\GeV$, and in cases where a massive b quark 
		is defined, the mass is set to $\mbottom = 
		4.75\GeV$.
		The transverse mass is defined as $\smash{\mTi = \sqrt{m_i^2 + \pTi^2}}$ and used to calculate the 
		scalar sum $\HT = \sumttbbg 
		\mTi$.
		The \POWHEG specific \hdamp value is given for the first three generator setups, all others do 
		not use this parameter and are therefore marked with (\NA)~\cite{cmscollaboration2023inclusive}.\\}
	\begin{adjustbox}{width=\textwidth} 
	\label{tab:generatorsettings}
	\renewcommand{\arraystretch}{1.2}
	\begin{tabular}{lllllll}
	\multirow{2}{*}{Generator setup} & Process/ & Generator/ & 
	\multirow{2}{*}{Tune} & \multirow{2}{*}{PDF set} & \multirow{2}{*}{\hdamp} & 
	\multirow{2}{*}{Scales} \\
	& ME order & Shower & & & & \\
		\hline
    \textsc{Powheg+p8} & \ttbar/ &
	\POWHEG v2/ & \multirow{2}{*}{CP5} & 5FS NNPDF3.1 & \multirow{2}{*}{1.379\mtop} &
	\multirow{2}{*}{$\muF = \muR = \mTtop$} \\
	\ttbar \textsc{5fs} & NLO & \PYTHIA 8.240 & & NNLO & & \\
	\textsc{Powheg+h7} & \ttbar/ &
	\POWHEG v2/ & \multirow{2}{*}{CH3} & 5FS NNPDF3.1 & \multirow{2}{*}{1.379\mtop} &
	\multirow{2}{*}{$\muF = \muR = \mTtop$} \\
	\ttbar \textsc{5fs} & NLO & \HERWIG 7.13 & & NNLO & & \\
	\textsc{Powheg+ol+p8} & \ttbb/ &
	\POWHEGBOX/ & \multirow{2}{*}{CP5} & 4FS NNPDF3.1 & \multirow{2}{*}{1.379\mtop} &
	$\smash{\muR=\frac12 \prodttbb \mTi^{1/4}}$, \\
	\ttbb \textsc{4fs} & NLO & \PYTHIA 8.240 & & NNLO as 0118 & &
	$\muF=\HT/4$ \\
	\textsc{sherpa+ol}& \ttbb/ & \multirow{2}{*}{\SHERPA 2.2.4} &
	\multirow{2}{*}{\SHERPA} & 4FS NNPDF3.0 & \multirow{2}{*}{\NA} &
	$\muR = \prodttbb \mTi^{1/4}$, \\
	\ttbb \textsc{4fs}& NLO & & & NNLO as 0118 & &
	$\muF = \HT/2$ \\
	\textsc{mg5\_}a\textsc{mc+p8} & \ttbb/ &
	\MGvATNLO & \multirow{2}{*}{CP5} & 4FS NNPDF3.1 & \multirow{2}{*}{\NA} &
	\multirow{2}{*}{$\muF = \muR = \sum \mT$} \\
	\ttbb \textsc{4fs} & NLO & v2.4.2/\PYTHIA 8.230 & & NNLO as 0118 & & \\
	\multirow{2}{*}{\textsc{mg5\_}a\textsc{mc+p8}} & \multirow{2}{*}{\ttjets 
	FxFx/} &
	\multirow{2}{*}{\MGvATNLO} & \multirow{3}{*}{CP5} & \multirow{2}{*}{5FS NNPDF3.1} & 
	\multirow{3}{*}{\NA} &
	$\muF = \muR = \sum \mT$, \\
	\multirow{2}{*}{\ttjets \textsc{FxFx 5fs}} & \multirow{2}{*}{NLO [$\leq$2 jets]} & 
	\multirow{2}{*}{v2.6.1/\PYTHIA 8.240} & & 
	\multirow{2}{*}{NNLO} & &
	$\textrm{qCut}=40\GeV$, \\
	& & & & & & $\textrm{qCutME}=20\GeV$ \\
	\end{tabular}
	\end{adjustbox}
\end{table}

\section{Phase space regions and observables}
Four different phase space regions are defined, which specifically examine different facets of \ttbb production.
These are partially overlapping and therefore not orthogonal.
The first region requires at least five jets with at least three of them being b jets (``\fjtbLONG'').
This is the most inclusive of all the regions considered and takes into account that one of the 
expected jets may be located outside the acceptance region.
A second region, labeled as ``\sjfbLONG'', targets fully resolved \ttbb events and requires at least six jets of which at least four must be b jets.
Two additional regions focus on the additional radiation of further light-flavour (i.e.\ non b) jets, these 
demand at least six (seven) jets, at least three (four) b jets and at least three 
light-flavour jets and are labeled as ``\sjtbtlLONG'' (``\sejfbtlLONG'').
Each of these four fiducial phase space regions additionally requires exactly one electron or muon.
Jets considered must meet $\pT > \SI{30}{GeV}$ and $\left|\eta\right| < 2.4$ thresholds.
All detailed specifications for object and event reconstruction, particle level definitions as well as 
event selection can be found in~\cite{cmscollaboration2023inclusive}.
Together with the inclusive cross section for all four regions previously described, a total of 37 different observables are defined.
Four of these normalized differential cross section distributions are discussed in Section~\ref{sec:normdiffxs}.

\section{Normalized differential cross sections\label{sec:normdiffxs}}
\begin{figure}[!h!tbp]
	\centering
    \begin{tabular}{@{}c@{}c@{}}
	\includegraphics[width=0.49\textwidth]{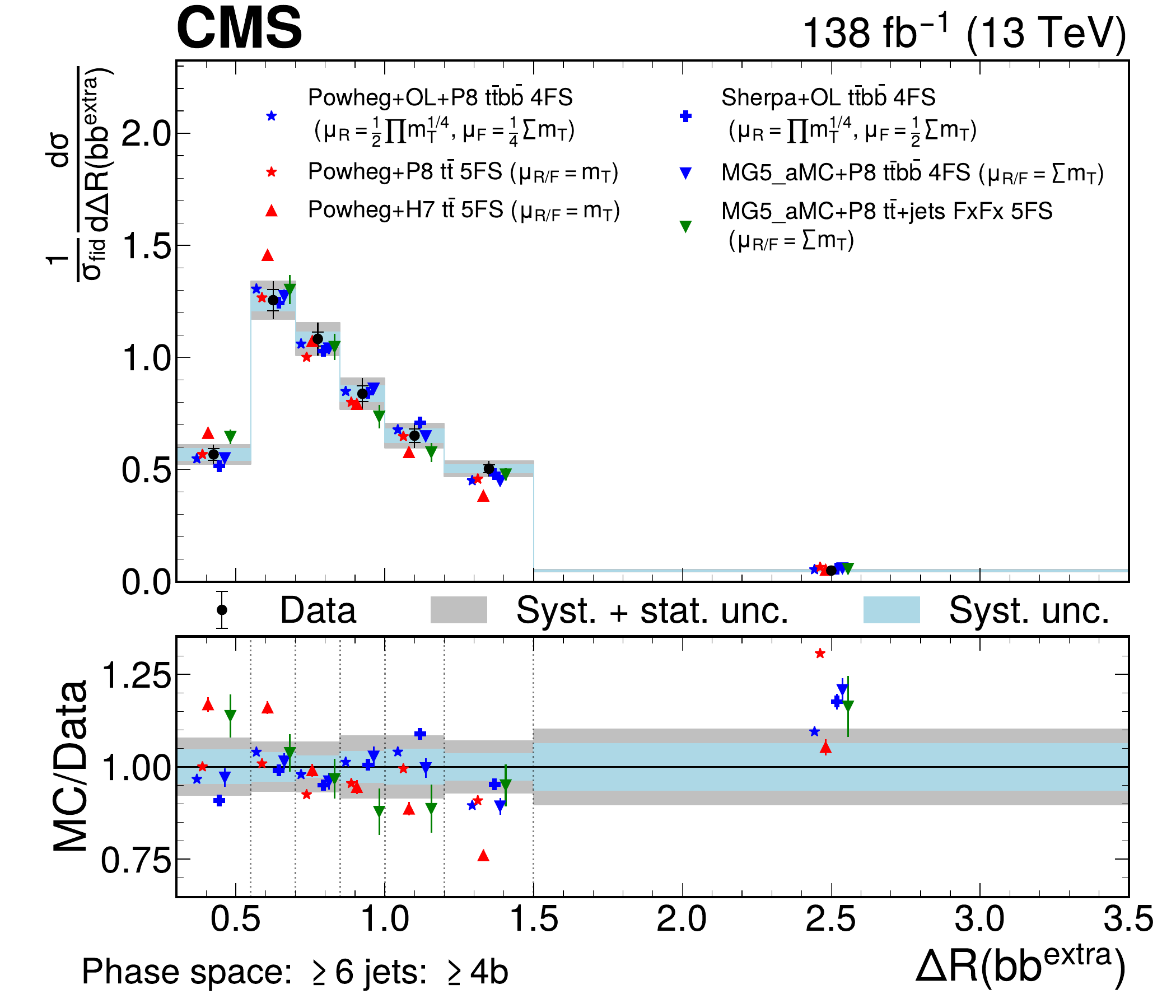} &
	\includegraphics[width=0.49\textwidth]{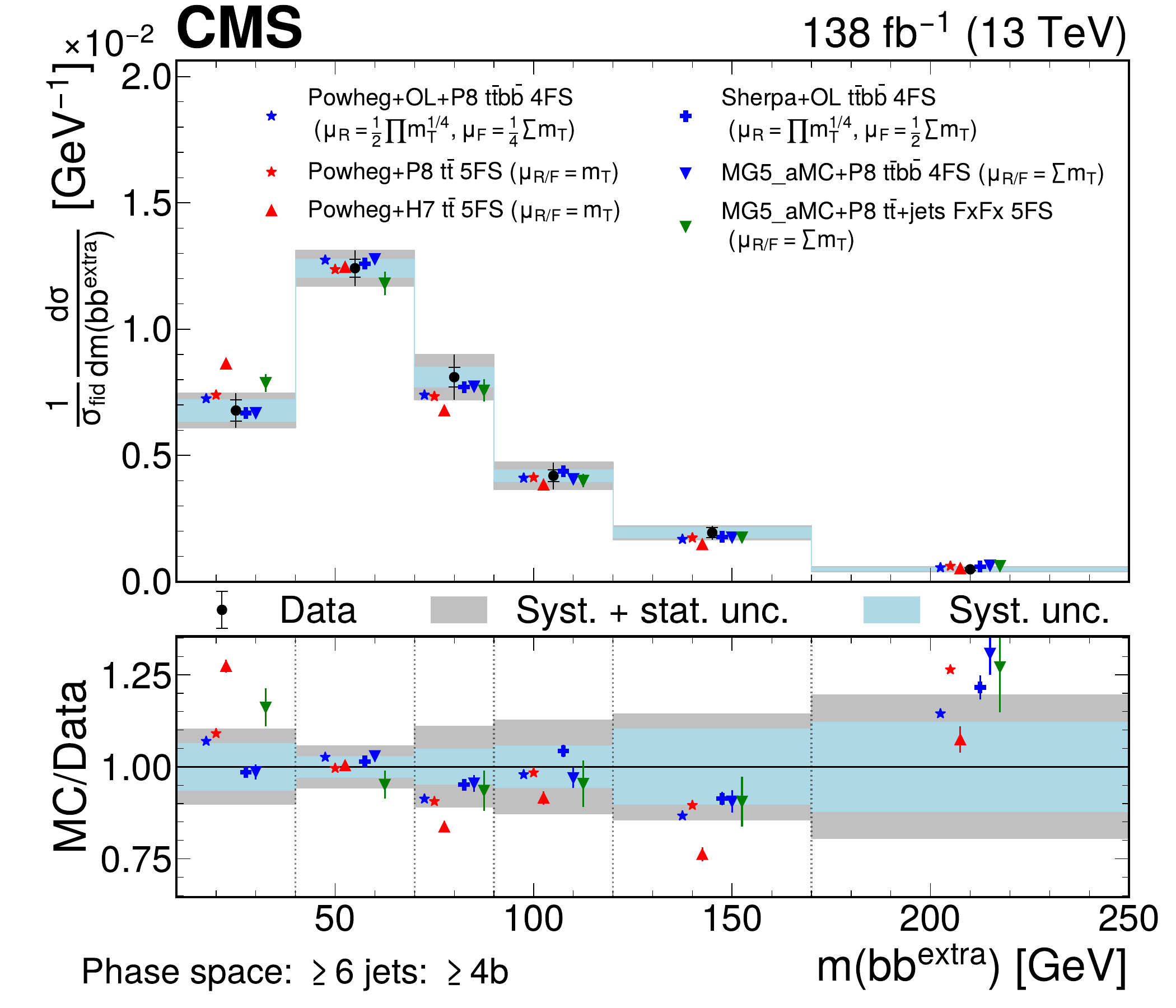} \\
	\includegraphics[width=0.49\textwidth]{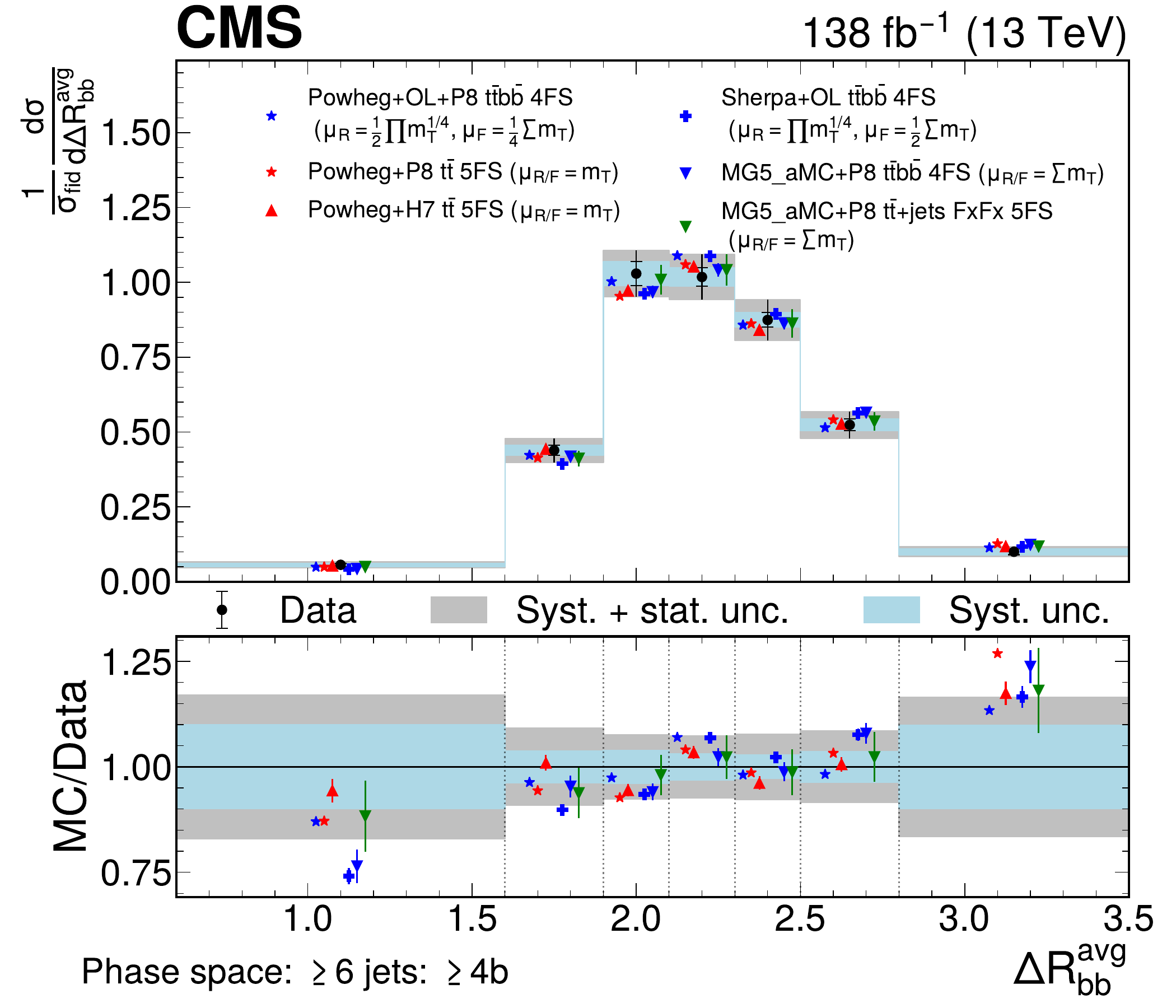} &
	\includegraphics[width=0.49\textwidth]{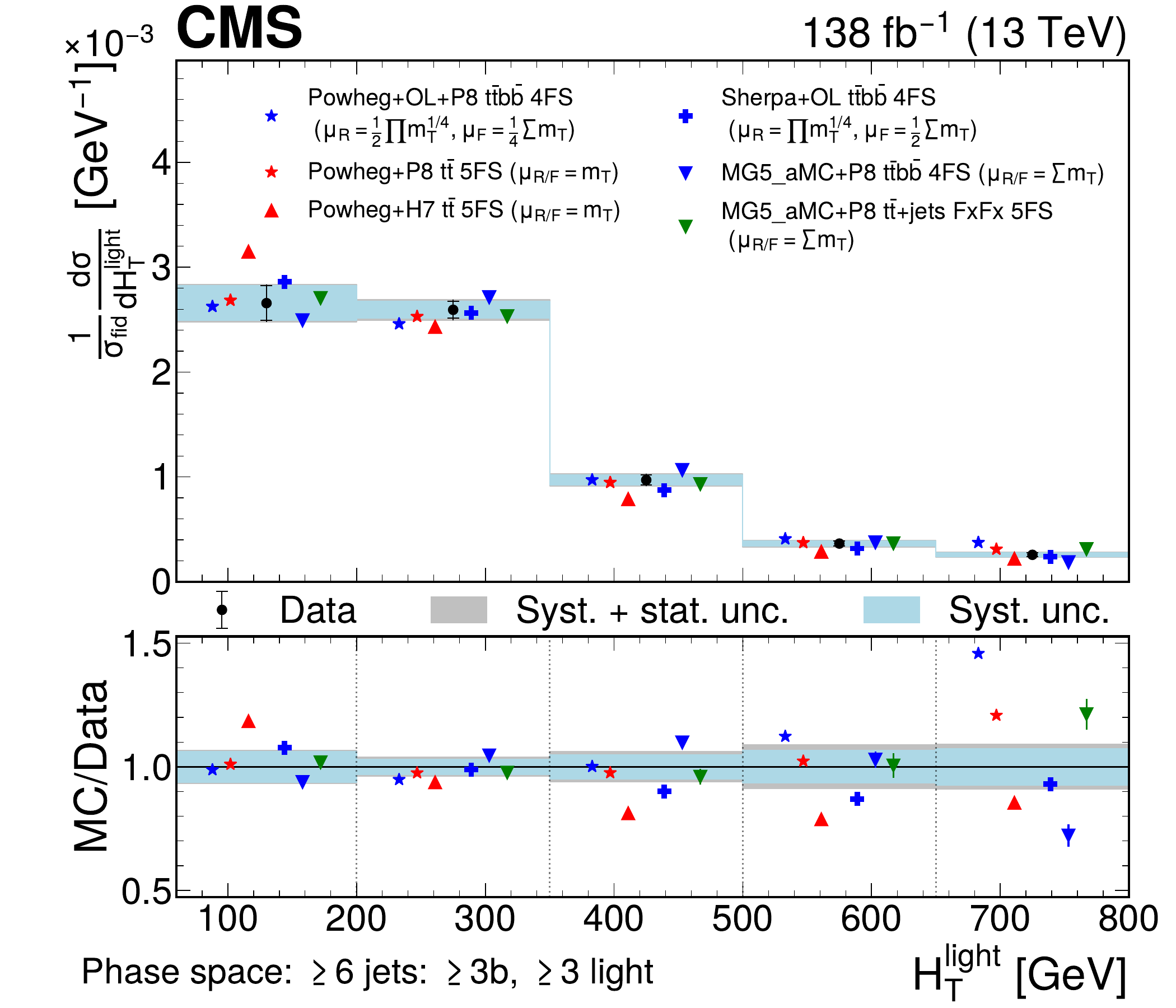}
	\end{tabular}
	\caption{Predicted and observed normalized differential cross sections in the \sjfbLONG fiducial phase space, for the \deltaR (upper left), invariant mass (upper 
	right) of the extra b-jet pair (\bbextra), average \dRbbavg of all b jets in an event (lower left), and the 
	scalar \pT sum of all light jets \HTl in the \sjtbtlLONG phase space. Data 
	represented by points. Inner (outer) vertical bars as well as blue (grey) bands are indicating the 
	systematic (total) uncertainties. The predictions include their statistical uncertainties due to limited 
	number of simulated events~\cite{cmscollaboration2023inclusive}.}
	\label{fig:diff}
\end{figure}

Two of the observables in the ``\sjfbLONG'' region discussed below are based on the definition of 
the pair of b jets which have the smallest spatial separation in 
the $\left(\eta,\phi\right)$-plane.
For this, every possible b jet combination is calculated according to
\begin{equation}
	\dRbb = \sqrt{ (\Delta \eta_{\bb})^2 + (\Delta \phi_{\bb})^2 } \qquad.
\end{equation}
The b jet pair with the smallest spatial distance \dRbb is then labeled as \bbextra in the following.
With this methodology, the b~jets determined in this manner are the two additional b~jets that do not 
originate from the top quark decay in 49\% of the events in the 
phase space under scrutiny~\cite{cmscollaboration2023inclusive}.
In this way, observables with properties expected to be sensitive to the modeling of the gluon splittings to \bb are considered.
The two observables defined on this basis are the spatial distance (\dRbbextra) and the invariant 
mass of the \bb pair (\mbbextra).
The distributions of these observables are shown in Figure~\ref{fig:diff}.
It can be seen that the distribution of the spatial distance \dRbbextra for the simulations \ttbbPP and 
\ttbbSherpa agree quite well with data in this 
observable.
In contrast, the modeling approach \ttbarPP shows distances \deltaR that are too large.
This cannot simply be attributed to the different calculation at ME level, as the \ttbarPH model also 
shows a deviation, but with an opposite trend towards too small distances \dRbbextra.
This observable is mainly driven by the PS, which is different in both \ttbar modeling approaches.
This indicates that the different PS configurations are not optimally tuned in this observable.
The deviating behavior of the event simulation in this crucial observable can also be seen in the \ttbb 
modeling performed in the context of the LHC Higgs Working Group~\cite{ferencz2023study}.
The behavior of the distribution of the invariant mass of the \bbextra pair is similar to the distance \dRbbextra. 
Here, the simulations \ttbbPP and \ttbbSherpa describe the invariant mass quite well compared to the measured data, while the modeling approach \ttbarPH tends to 
predict invariant masses that are too small.

Another interesting observable is the average spatial distance \deltaR of all b jets in an event,  
\dRbbavg. 
Such an observable is used, for example, in \ttHbb analyses as an helpful input feature for artificial neural networks in the separation of signal and 
backgrounds~\cite{CMS:2020grm}.
In Figure~\ref{fig:diff} (lower left) it can be seen that all modeling approaches examined tend to deliver 
on average \dRbbavg values that are too large compared to data.
However, it should be noted that the uncertainties of the measurement are still large, such that the 
trend is not (yet) very significant and mainly statistically limited.

An observable that probes the additional light-flavour jet radiation is the scalar \pT sum of all 
light-flavour jets in the \sjtbtlLONG phase space.
The normalized differential cross section is shown in Figure~\ref{fig:diff} (lower right).
This observable reveals that the prediction of \HTl in the \ttbbPP modeling approach is mis-modeled compared to the measured data since the \pT of the light jets 
tends to be predicted too large. 
On the other hand, in the \ttbarPH modeling approach the  \pT of the light-flavour jets is rather soft.
All other event simulations examined show a tendency to describe this observable more accurately.

\begin{figure}[!h!tbp]
	\centering
	\includegraphics[width=0.59\textwidth]{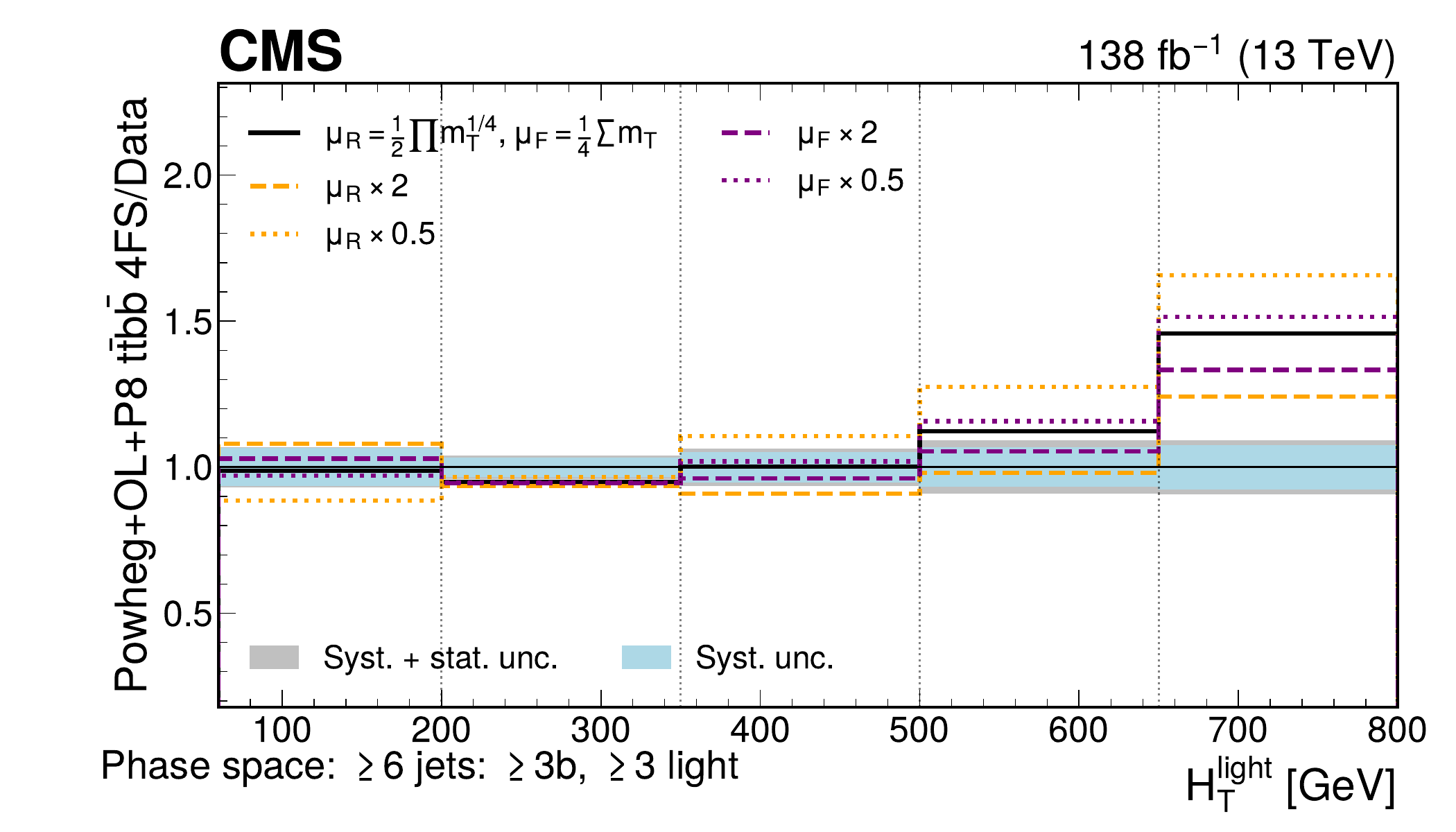}
	\caption{\ttbbPP modeling approach with varied \muR and \muF scale settings as a ratio of normalized differential cross section predictions to the measured 
		normalized differential cross sections for \HTl in the \sjtbtlLONG phase space. The systematic (total) uncertainties of the measurement are represented as blue 
		(grey) bands. Variations of the \muR (\muF) scale relative to the nominal scale setting are shown in orange (purple)~\cite{cmscollaboration2023inclusive}.}
	\label{fig:qcdscales_light}
\end{figure}

Important parameters for \ttbb simulations at the ME level are the renormalization and factorization 
scales $\muR$ and $\muF$. Various nominal models are used by the ATLAS and CMS Collaborations 
and are compared in~\cite{ferencz2023study}.
Therefore, it is important to check variations of $\muR,\muF$ and their effect on distributions 
depending on the choice of both scales. 
These scales are independently varied by a factor of 2 and 
0.5 and the corresponding re-weighted distributions of the 
\ttbbPP modeling approach are examined. This is shown for the \HTl observable in the \sjtbtlLONG  
phase space in Figure~\ref{fig:qcdscales_light}.

In the ratio of the predictions with varied scale settings to the measured data it can be observed that 
a larger choice of both scales would be preferred by the data in this particular observable.
This suggests the prediction of light-flavor jets with preferably large transverse momenta may not 
necessarily be a problem of the generator, instead it could be necessary to optimize the choice of 
scales being applied. Similar trends are observed for other observables, as discussed 
in~\cite{cmscollaboration2023inclusive}.

\section{Summary}
Normalized differential cross section measurements of the associated production of top 
quark-antiquark and bottom quark-antiquark pairs for events with precisely one
electron or muon have been presented.
These measurements are based on recorded data from proton-proton collisions by the CMS 
experiment at $\sqrt{s} = 
\SI{13}{TeV }$ and a corresponding to an integrated luminosity of 
$\SI{138}{fb^{-1}}$~\cite{cmscollaboration2023inclusive}.
Multiple observables are defined targeting b jets as well as additional light jet radiation produced in association with top quark pairs. 
The relative uncertainties range from $2\!-\!50\%$ depending on the phase space and 
the observable.
None of the modeling approaches examined simultaneously describe all measured differential distributions in all phase space regions. 
Thus, depending on the phase space and observable, mis-modeling  is still evident and optimization is necessary for a more accurate description of the observed 
processes.
A Rivet routine is available for tuning new modeling approaches~\cite{hepdata.138416.v1}.

\Acknowledgements
This research was supported by the German Federal Ministry of Education and Research (BMBF) 
under grant 05H21VKCCB. I would like to express my gratitude for BMBF's financial support.

\bibliography{eprint}{}
\bibliographystyle{unsrt}
 
\end{document}